\documentclass[10pt, conference]{ieeeconf}

\IEEEoverridecommandlockouts

\usepackage{graphicx}
\usepackage{amsmath}
\usepackage{amssymb}
\usepackage{cite}
\usepackage{xcolor}
\usepackage{tikz}
\usetikzlibrary{arrows.meta,positioning,fit,shapes,calc,backgrounds}
\usepackage{hyperref}
\hypersetup{hidelinks}
\usepackage{latexsym}
\usepackage{subfigure}
\usepackage{graphicx}
\usepackage{color}
\usepackage{amsmath}
\usepackage{mathtools}
\usepackage{booktabs}

\newcommand{\figref}[1]{Fig.~\ref{fig:#1}}

\newcommand{\tabref}[1]{Table~\ref{tab:#1}}

\renewcommand{\eqref}[1]{(\ref{eq:#1})}

\def\sign{\mathop{\rm sign}\nolimits}

\def\atan2{\mathop{\rm atan2}\nolimits}

\newtheorem{prop}{Proposition}

\def\atan2{\mathrm{atan2}}

\title{\LARGE \bf Event-Driven Safe Control of a Black-Box Balancing Wheel with Friction-Gated Actuation}

\author{Kanishka Harwani\thanks{K. Harwani and N. Navarkar are with NYU (email: k.H@nyu.edu, nsn4968@nyu.edu). A. Arab is with the Department of Mechanical and Aerospace Engineering at New York University, and also Grove School of Engineering at CCNY (email: aliasghar.arab@nyu.edu, aarab@ccny.cuny.edu). F. Khorrami is with the Department of Electrical and Computer Engineering at New York University (email: khorrami@nyu.edu).}, Neha Navarkar, Aliasghar Arab$^*$, Farshad Khorrami}

\begin{document}
\maketitle

\begin{abstract}
Commercial balancing wheels are closed systems: an internal controller keeps the board upright, and a rider steers by shifting weight and twisting the hips. We augment a one-wheel with a lead-screw weight shift for longitudinal motion and a reaction wheel for yaw. The reaction wheel stores limited momentum, so a turn is a sequence of torque pulses, and whether a pulse rotates the body depends on static friction at the tire contact. From a 24-pulse torque sweep on one surface we fit a two-part model that gives the probability that a pulse rotates the body, treating the breakaway torque as random, and the rotation angle when it does; at maximum torque the posterior engagement probability is 0.54 (90\% credible interval 0.24--0.84). The weight shift is modeled as a rate-limited input with a measured speed gain and an assumed first-order response, which predicts stopping distances four times those of a direct-force model. These models drive an event-triggered supervisor that schedules pulses by expected heading improvement, brakes when the predicted braking path would leave the safe set, tightens margins for model uncertainty, and updates the friction estimate online. In simulation on 500 plants drawn from the fitted models, the robust supervisor completed 92.6\% of missions with 1.8\% safety violations and 1.6\% safe aborts, versus 84.0\% and 7.2\% for a supervisor using mean models and 1.8\% and 20.4\% for a baseline that ignores friction. A video of the hardware experiments is available online (https://github.com/ASAS-Labs/x-wheel).
\end{abstract}

\section{Introduction}
\label{sec:intro}
Operating a new mobility tool safely and efficiently is a fundamental challenge in robotics and human-robot interaction. Professional riders master such devices through structured learning: they receive safety instructions, practice under guidance, and refine their skills while keeping safety as the top priority~\cite{karinch2000lessons}. Extreme athletes succeed at high-risk tasks through systematic practice that prioritizes safety while pushing performance boundaries~\cite{vande2021decision}, learning from failures and continuously refining their actions~\cite{gieseler2019learning}. Much of this skill concerns the body itself: a rider learns how much weight shift produces how much acceleration, and how sharply a hip twist turns the board on a given surface. This paper asks how an autonomous system can operate a black-box balancing wheel through the same body inputs, safely, when the response to those inputs is uncertain.

\begin{figure}[t!]
\centering
\begin{tikzpicture}
  \node[anchor=south west, inner sep=0] (image) at (0,0)
    {\includegraphics[width=3.3in]{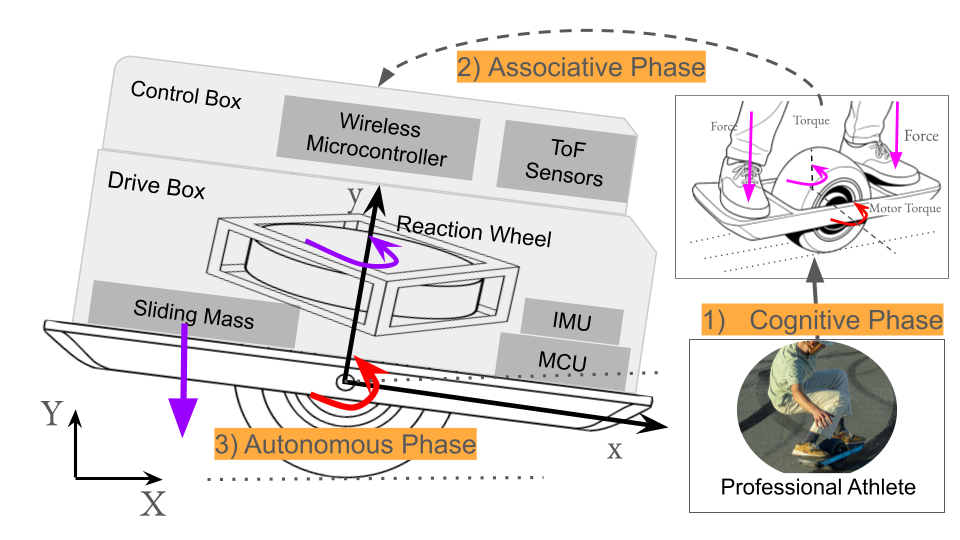}};
  \node[anchor=south west, font=\small, color=black,
        fill=white, fill opacity=0.3, text opacity=1]
        at ($ (image.south west) + (2.7cm,2.4cm) $) {$\tau_r$};
  \node[anchor=south west, font=\small, color=black,
        fill=white, fill opacity=0.7, text opacity=1]
        at ($ (image.south west) + (2.4cm,1.3cm) $) {$\tau_w$};
  \node[anchor=south west, font=\small, color=black,
        fill=white, fill opacity=0.2, text opacity=1]
        at ($ (image.south west) + (1.3cm,0.5cm) $) {$mg$};
  \node[anchor=south west, font=\large, color=black,
        fill=white, fill opacity=0.5, text opacity=1]
        at ($ (image.south west) + (3.9cm,1.1cm) $) {$\theta$};
\end{tikzpicture}
\caption{Balancing wheel with human-inspired actuation. A sliding mass reproduces rider weight shift and a reaction wheel (torque $\tau_r$) reproduces hip twist; $\theta$ is the deck pitch set by the internal balance controller.}
\label{fig:architecture}
\vspace{-3mm}
\end{figure}

Personal balancing wheels are designed to be ridden, not driven. A one-wheel maintains pitch balance with an internal controller whose structure and gains are not exposed, and a rider steers by shifting weight and twisting the hips. We keep the commercial balance loop intact and add a module that reproduces these rider actions: a lead-screw weight shift for longitudinal acceleration and a deck-mounted reaction wheel that imparts yaw momentum like a hip twist (\figref{architecture}). Our approach combines three elements: (i) human-inspired actuation that acts on the black box only through rider-like inputs, (ii) data-driven models of each actuator with explicit uncertainty, and (iii) safety instructions, such as keeping distance from a person, encoded as constraints in a supervisor that accounts for that uncertainty.

Both actuators are uncertain, in different ways. The reaction wheel exchanges momentum with the body, so the momentum per pulse is bounded by the flywheel inertia and speed limit, and the flywheel must spin down before the next command. A turn is therefore a sequence of pulses, and whether a pulse rotates the body at all depends on static friction at the tire contact patch~\cite{armstrong1994survey,canudas1995new}, which is not known in advance. In separate heading-command trials on other surfaces, 46\% of 24 turns showed measurable rotation. The weight shift is smoother but slew-limited, and the black-box balance loop filters its effect on speed, so stopping takes longer than a direct-force model predicts.

Learning and expert guidance can accelerate skill acquisition: instructed learning~\cite{arab2021instructed_acc} shows how guidance shapes what is learned, and agentic control~\cite{azdam2025maneuvergpt} shows how a system can propose professional-level maneuvers. Data-driven models with calibrated uncertainty are central to this process. Gaussian Processes (GPs) provide such uncertainty from small datasets~\cite{rasmussen2006gaussian_acc} and have been used in learning-based and event-triggered control~\cite{umlauft2017feedback_acc,hewing2020learning,umlauft2019feedback_acc}. Existing approaches, however, often treat learning and safety separately, which leads either to conservative behavior or to violations while the model is still inaccurate.

On the safety side, Control Barrier Functions (CBFs) ensure forward invariance of safe sets~\cite{ames2019control_acc}, and their integration with Nonlinear Model Predictive Control (NMPC) has become a standard paradigm for safety-critical systems~\cite{schilliger2021control_acc,arab2021safe_acc,liu2025motion}. Hierarchical CBF prioritization encodes safety instructions for human-robot interaction~\cite{maithani2025proactive_acc,arab2024ccta}, and probabilistic CBFs address learned dynamics~\cite{khojasteh2020probabilistic}. NMPC has been extended with stability guarantees for avoidance~\cite{santos2024stability_acc}, online adaptation~\cite{gaber2024adaptive}, learned hierarchical models~\cite{zhou2024learning}, and data-driven terminal conditions~\cite{worthmann2024data_acc}, while stochastic and robust predictive control handle bounded or probabilistic disturbances~\cite{mesbah2016stochastic,mayne2005robust}. These methods assume that the commanded input is realized, possibly with bounded error. A friction-limited reaction wheel violates this assumption: its response is a discrete engagement event rather than a continuous gain. Control over unreliable actuation channels~\cite{imer2006optimal,schenato2007foundations} models such losses as independent drops, and event-triggered control~\cite{tabuada2007event,heemels2012introduction_acc,khalili2026global} updates commands only when state-dependent conditions fire, which fits impulsive, dwell-limited actuation.

The yaw data are few and binary, so we use standard small-sample tools. A hurdle model~\cite{mullahy1986specification} splits a response into a binary event (does the body rotate?) and a magnitude given that it does. The binary part is a probit model: if the breakaway torque varies normally between pulses, the engagement probability is the normal cumulative distribution function of the standardized torque. The bootstrap~\cite{efron1993introduction} estimates parameter uncertainty by refitting resampled data, and Wilson score intervals~\cite{wilson1927probable} give confidence intervals for proportions that remain reliable for small samples. In simulation, all controllers share the same random plants and noise, so their differences are paired and tested with McNemar's test~\cite{mcnemar1947note}.

This paper connects these elements for a black-box balancing wheel. Its contributions are: (1) human-inspired actuation of a closed commercial one-wheel through a weight shift and a momentum-limited reaction wheel, with a characterization of friction-gated yaw engagement; (2) actuator models with explicit uncertainty, namely a data-driven hurdle model (probit breakaway threshold and lognormal rotation magnitude) for yaw and a slew-limited model with a measured gain and an assumed first-order lag for the longitudinal channel; (3) an uncertainty-aware hybrid supervisor with a backup-safe braking policy, using expected-improvement pulse scheduling, a safety check over the full braking path, uncertainty-tightened margins with a per-event probabilistic bound, and an online friction posterior with safe abort; and (4) a Monte Carlo evaluation with baselines, ablations, parameter mismatch, and plants whose structure departs from the fitted models.

Section~\ref{sec:platform} describes the platform and actuation, Section~\ref{sec:learning} the actuator models, Section~\ref{sec:control} the supervisor, and Section~\ref{sec:results} the simulation results; Section~\ref{sec:conclusions} concludes the paper.

\section{Platform and Human-Inspired Actuation}
\label{sec:platform}
\begin{figure}[t!]
\centering
\resizebox{\columnwidth}{!}{%
\begin{tikzpicture}[x=1mm, y=1mm,
  font=\scriptsize,
  blk/.style={draw, rounded corners=2pt, minimum width=30mm,
              minimum height=7mm, align=center, inner sep=1.5pt,
              line width=0.3pt},
  sense/.style={blk, fill=black!5},
  comp/.style={blk, fill=blue!8},
  tag/.style={draw, rounded corners=1.5pt, align=center, inner sep=1.2pt,
              line width=0.3pt, fill=white, fill opacity=0.85, text opacity=1,
              font=\tiny},
  act/.style={tag, fill=orange!14},
  bb/.style={tag, fill=yellow!25},
  ar/.style={-{Latex[length=1.5mm, width=1.1mm]}, line width=0.35pt},
  ld/.style={line width=0.3pt, {Circle[length=0.9mm]}-},
]
\node[anchor=south west, inner sep=0] (photo) at (34,0)
  {\includegraphics[height=46mm]{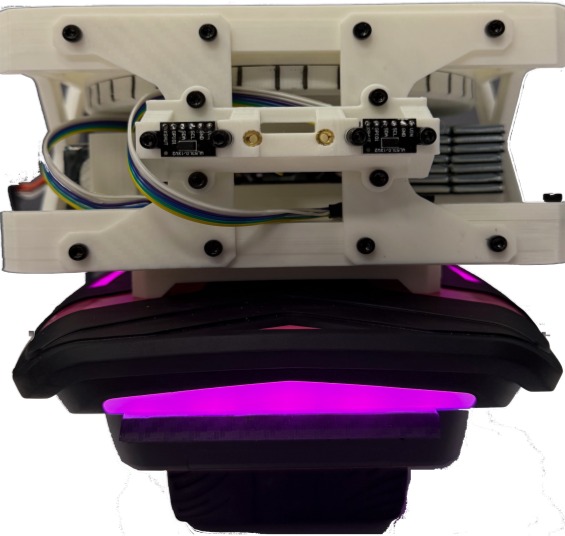}};
\node[sense] (cam) at (16,40) {Overhead camera\\[-1pt]\tiny AprilTag pose, 30\,Hz};
\node[sense] (est) at (16,28) {State estimate\\[-1pt]\tiny pose, velocity};
\node[comp]  (sup) at (16,14) {Event-triggered supervisor\\[-1pt]\tiny learned actuator models};
\draw[ar] (cam) -- (est);
\draw[ar] (est) -- (sup);
\node[tag, fill=blue!8] (esp) at (50,43.3) {ESP32-S3 (TCP)};
\draw[ar] (sup.east) -- ++(2.5,0) |- (esp.west);
\node[act, anchor=west] (rw) at (71,37) {Reaction wheel (twist)\\[-1pt]$\tau_{\max}=0.123$\,N$\cdot$m};
\draw[ld] (62,39.6) -- (rw.west);
\node[act, anchor=west] (ws) at (71,27) {Weight shift (lean)\\[-1pt]$\bar{r}=11.5$ counts/s};
\draw[ld] (66,28) -- (ws.west);
\node[tag, anchor=east] (tof) at (44,29) {ToF};
\draw[ld] (50,34.3) -- (tof.east);
\node[bb, anchor=west] (bbc) at (71,11) {Black-box balance\\[-1pt]controller};
\draw[ld] (66,15) -- (bbc.west);
\draw[ar, dashed] (38.5,40) -- (cam.east);
\end{tikzpicture}}
\vspace{-5mm}
\caption{Platform and signal path. The supervisor commands the weight shift and reaction wheel through the ESP32-S3; the internal balance controller is not commanded.}
\label{fig:system}
\vspace{-2mm}
\end{figure}

The platform is a commercial single-wheel board with an unmodeled internal balance controller, augmented with two actuators that reproduce rider weight shift and hip twist (\figref{architecture}); the signal path is shown in \figref{system}. Fore-aft weight shift uses a geared DC motor on a lead screw with potentiometer feedback. Yaw torque uses a reaction wheel through an ODrive Micro on CAN. An ESP32-S3 sends setpoints over TCP; an overhead camera tracks a tilted \texttt{tag36h11} fiducial at 30\,Hz. Five VL53L0X time-of-flight (ToF) sensors on the control box, facing front, front-left, front-right, back-left, and back-right, measure range to surrounding surfaces up to 1.8\,m together with filtered range rates; they are logged but not used by the supervisor in this study. Parameters are summarized in \tabref{params} (measured, learned, assumed, or design).

Here the hardware supplies all identification data (torque sweep, weight-shift slew and gain, translating trials), with camera pose labeling displacement and rotation; the supervisor runs only in closed-loop simulation, and \figref{system} shows its intended hardware signal path.

\begin{table}[t!]
\caption{Actuator parameters (M measured, L learned, A assumed, D design).}
\label{tab:params}
\centering
\footnotesize
\setlength{\tabcolsep}{2.5pt}
\begin{tabular}{@{}lcc@{}}
\hline
Parameter & Value & Src. \\
\hline
$\tau_{\max}$ & 0.123 N$\cdot$m & M \\
Flywheel limit & 100 (38 peak) rev/s & M \\
$T_b$, $T_d$ & 0.8 s, 3.3 s & D/M \\
$\mu_s$, $\sigma_s$ & 0.117, 0.018 N$\cdot$m & L \\
$|\Delta\psi|$, med.\ (90\% pred.) & $26^\circ$ ($7$--$98^\circ$) & L \\
Motion thresh., latency & 17 deg/s, 0.75 s & L \\
Slew $\bar{r}$ & 11.5 counts/s & M \\
Shift range (neutral) & 10--240 (130) ct & M \\
$k_v$ & 17.7 mm/s/ct & M \\
$T_v$ (balance lag) & 0.8 s & A \\
$k_v$, $T_v$ uncert. & $\pm 20\%$, $\pm 30\%$ & A \\
$\sigma_w$ & 15 deg/$\sqrt{\mathrm{m}}$ & L$^*$ \\
$\epsilon$, $p_{\min}$, $m$ & $1^\circ$, 0.1, 0.1 m & D \\
\hline
\multicolumn{3}{@{}l@{}}{\footnotesize $^*$translating trials; systematic, not clean drift.}
\end{tabular}
\vspace{-3mm}
\end{table}

\subsection{Rotational Actuation by Reaction Momentum}
Let $\psi$ and $\omega_b$ denote body heading and yaw rate, $I_b$ the body yaw inertia, and $J_r$ and $\omega_r$ the inertia and angular rate of the flywheel about its vertical spin axis. A motor torque $\tau$ between flywheel and body gives
\begin{equation}
\label{eq:yaw}
J_r\dot{\omega}_r = \tau, \qquad I_b\dot{\omega}_b = -\tau - \tau_f,
\end{equation}
\noindent with yaw friction torque $\tau_f$ at the tire contact. With static breakaway torque $\tau_s$ and kinetic friction torque $\tau_k \le \tau_s$, the contact sticks ($\omega_b = 0$, $\tau_f = -\tau$) while $|\tau| \le \tau_s$; otherwise it slips with $\tau_f = \tau_k\sign(\omega_b)$~\cite{armstrong1994survey}. The flywheel momentum $|J_r\omega_r|$ is bounded, so a turn is a pulse of duration $T_b$; spin-down before the next pulse requires $|\tau| < \tau_s$, imposing dwell $T_d$. On our platform $\tau_s$ is comparable to $\tau_{\max}$; at the observed 100\,rev/s flywheel limit and $T_b=0.8$\,s the momentum bound is not active (38\,rev/s peak observed). Pulse-duration modulation remains unexplored. Breakaway torque varies with surface; a pulse either rotates the body opposite to the flywheel torque or does not.

\subsection{Longitudinal Actuation by Weight Shift}
Let $\delta$ be lead-screw position in counts from neutral and $v$ forward speed. We model
\begin{equation}
\label{eq:long}
\dot{\delta} = \mathrm{sat}_{\bar{r}}(\delta_c - \delta), \qquad T_v\dot{v} = k_v\delta - v,
\end{equation}
\noindent with command $\delta_c$, measured slew limit $\bar{r}$, measured gain $k_v$, and assumed lag $T_v$. Shifting the mass by $\delta$ moves the center of mass ahead of the axle; the internal controller rebalances by pitching the deck by $\theta$ (\figref{architecture}) and driving forward. Since $\theta$ is not observed, \eqref{long} maps $\delta$ directly to $v$, and $T_v$ lumps the pitch and speed transients. Braking uses $\delta_c = -\mathrm{sat}_{25}(2v/k_v)$ with nominal $k_v$. Reversing $\delta$ over 50 counts takes 4.3\,s.

\section{Learning the Actuator Models}
\label{sec:learning}
The yaw response is discontinuous because of the friction threshold, so we use a hurdle model that separates engagement from magnitude. The longitudinal channel uses the low-order model \eqref{long}.

\subsection{Torque-Sweep Data}
We commanded 24 pulses of duration $T_b = 0.8$\,s at six torque levels on talc-treated wood, in two sets of 12, each consisting of two ascending passes over the levels with alternating sign (talc applied once; location varied). At 0.11 and 0.123\,N$\cdot$m the sign is therefore fixed by the pass, so order and sign effects cannot be separated. We label engagement when the operator marked rotation opposite to the flywheel torque. Net planar displacement before versus after each trial exceeded 0.05\,m in five cases, which were excluded, leaving 19 stationary trials; three of the five showed heading change with the sign of the commanded torque, consistent with path curvature rather than reaction torque. Two retained trials had displacement 0.039--0.040\,m.

\subsection{Engagement Model}
\begin{figure}[t!]
\centering
\includegraphics[width=\columnwidth]{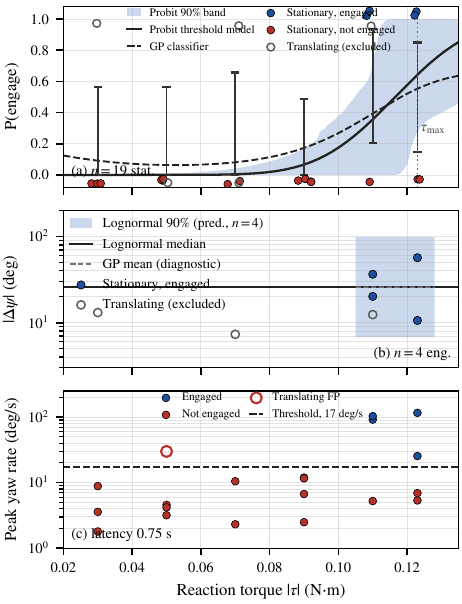}
\vspace{-6mm}
\caption{Reaction-wheel model. (a) Engagement with per-level Wilson intervals ($n=19$ stationary). (b) Engaged magnitude: lognormal predictive band ($n=4$); GP shown as diagnostic. (c) Peak yaw rate and first-motion detector ($n=19$).}
\label{fig:yaw}
\vspace{-4mm}
\end{figure}

Contact conditions vary between pulses, so we treat the breakaway torque of each pulse as an independent draw $\tau_s \sim \mathcal{N}(\mu_s, \sigma_s^2)$. By \eqref{yaw} the body rotates when $|\tau| > \tau_s$, hence
\begin{equation}
\label{eq:probit}
P(\text{engage} \mid \tau) = \Pr[\tau_s < |\tau|] = \Phi\!\left(\frac{|\tau| - \mu_s}{\sigma_s}\right),
\end{equation}
\noindent where $\Phi$ is the standard normal cumulative distribution function. \figref{yaw}(a) shows per-level empirical rates with Wilson 95\% intervals: 0/12 below 0.09\,N$\cdot$m; 2/3 at 0.11 and 2/4 at 0.123\,N$\cdot$m. A grid posterior with a weak log-normal prior on $\sigma_s$ remains proper when torque separates the outcomes and gives $\mu_s = 0.125$\,N$\cdot$m (90\% credible interval 0.104--0.161), $\sigma_s = 0.025$\,N$\cdot$m (0.011--0.073), and $P(\text{engage}\mid\tau_{\max}) = 0.54$ (0.24--0.84); a GP classifier is consistent at 0.55. The supervisor uses the maximum-likelihood fit $\mu_s = 0.117$\,N$\cdot$m, $\sigma_s = 0.018$\,N$\cdot$m, and $P(\text{engage}\mid\tau_{\max}) = 0.64$, with a 2000-sample bootstrap (90\% interval of $\mu_s$ 0.105--0.138) to generate plants. The ordered design makes the pulses non-exchangeable, and in 17\% of resamples torque separates the outcomes and $\sigma_s$ is clipped to 0.002\,N$\cdot$m, so the bootstrap is a plant generator rather than a calibrated interval, and Section~\ref{sec:results} repeats the evaluation with plants drawn from the posterior. The posterior predictive count at every torque level contains the observed count within its 90\% interval. Fitted to one set, the model predicts the other with log score $-3.0$ and $-3.3$, against $-5.1$ and $-4.8$ for a torque-independent rate, and predicts 1.8 and 2.3 engagements against 2 observed in each. A set offset in breakaway torque is estimated at 0.004\,N$\cdot$m ($-0.046$ to 0.051), whereas a second-pass offset of 0.027\,N$\cdot$m ($-0.015$ to 0.069) cannot be excluded and is confounded with sign; Section~\ref{sec:results} therefore tests a sign-dependent breakaway as an out-of-model plant. A hierarchical model $\tau_{s,ij} = \mu_s + b_i + \epsilon_{ij}$ with session effects $b_i$ requires sessions on several surfaces. $\mu_s$ varies by at most 0.002\,N$\cdot$m over stationarity cutoffs 0.03--0.10\,m.

\subsection{Rotation Magnitude and Engagement Detection}
All four stationary engagements rotated opposite to the flywheel torque. Magnitudes $11^\circ$--$57^\circ$ are modeled as lognormal with median $26^\circ$ and 90\% predictive range $7$--$98^\circ$ on $n=4$, which widens to $4$--$175^\circ$ under the Student-t predictive that also accounts for uncertainty in the log spread; a GP on $\ln|\Delta\psi|$ assigns variation to noise and does not resolve torque dependence over 0.11--0.123\,N$\cdot$m. Engaged onset times cluster at 0.73--0.78\,s (near pulse end), suggesting ramp/latency effects. Peak yaw rate separates outcomes; threshold 17\,deg/s is the geometric mean of class extrema. First-motion latency averages 0.75\,s on engaged trials; one translating non-engaged trial reached 29.9\,deg/s (in-sample margin). The detector label uses operator engagement marking consistent with reaction direction.

\subsection{Longitudinal Model}
\begin{figure}[t!]
\centering
\includegraphics[width=\columnwidth]{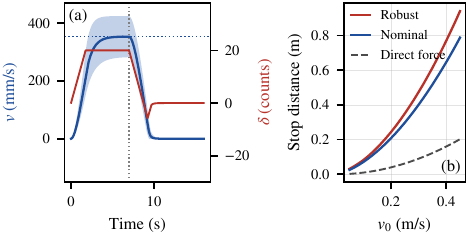}
\vspace{-6mm}
\caption{Weight-shift model ($T_v$ assumed). (a) Speed band over the uncertainty box and $\delta$ for cruise then brake at 7\,s. (b) Stopping distance versus $v_0$.}
\label{fig:long}
\vspace{-4mm}
\end{figure}

Only two longitudinal quantities are identified from data: $\bar{r} = 11.5$\,counts/s and the steady-state gain $k_v = 17.7$\,mm/s per count at $+20$ counts. The lag $T_v = 0.8$\,s and the box ($\pm 20\%$ on $k_v$, $\pm 30\%$ on $T_v$) are assumed, so this model is an assumption anchored by one measured gain rather than a data-driven model; Section~\ref{sec:results} therefore varies $T_v$, $k_v$, and the model structure. At 354\,mm/s the model predicts stopping distance 0.53\,m (0.63\,m robust corner), versus 0.13\,m for a direct 0.5\,m/s$^2$ force model (\figref{long}). $\sigma_w = 15$\,deg/$\sqrt{\mathrm{m}}$ is estimated from five translating trials whose heading change matched commanded-torque sign (systematic, not isolated drift).

\section{Hybrid Supervisory Control with Backup-Safe Braking}
\label{sec:control}
\begin{figure}[t!]
\centering
\includegraphics[width=\columnwidth]{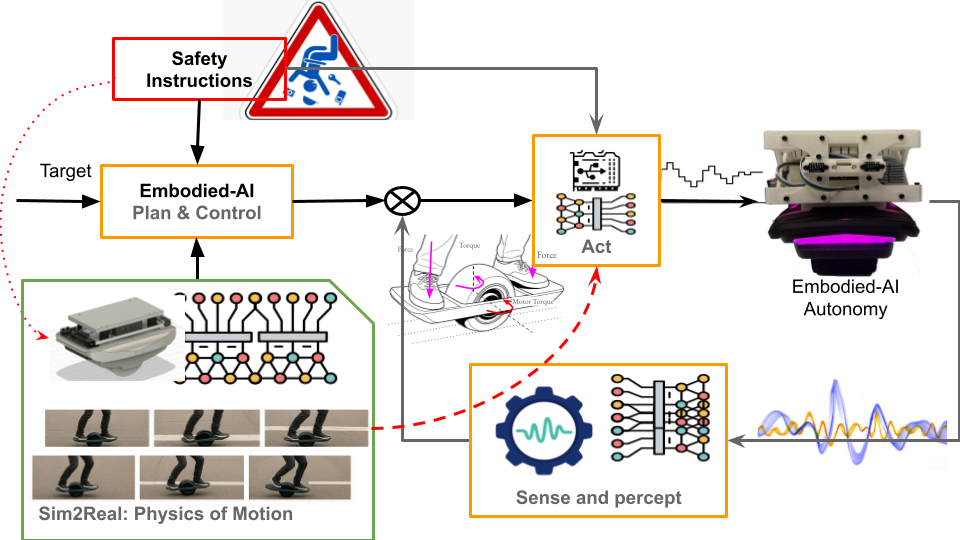}
\caption{Control architecture. Sim2Real denotes the hardware identification data from which the actuator models and the simulation plant are built; Sense and percept denotes camera pose. Here the loop is closed in simulation.}
\label{fig:control_arch}
\vspace{-3mm}
\end{figure}

Actuator limits make event-driven supervision natural: commands update when triggers fire at the camera rate. \figref{control_arch} shows the resulting loop. The supervisor alternates stopped mode (yaw pulses after dwell $T_d$) and driving mode (longitudinal cruise, no yaw). While driving it commands $\delta_c = 20$ counts, reduced to $\max(8, 20 d/1.2)$ counts for short hops of distance $d$. Waypoints register only when stopped within 0.35\,m. Yaw pulses apply over a 1\,s rotation profile; the detour planner replaces a waypoint by a tangent point at keep-out radius $\rho_{\mathrm{plan}}+0.45$\,m when the segment to the goal cuts the inflated disk. Arena margins use the same fixed $m$ as \eqref{margin}. We assume a static, known person location; after a safe stop the mission remains feasible along the detour.

\subsection{Problem Statement and Backup-Safe Supervision}
The yaw input is a discrete engagement event followed by a dwell $T_d$, so at each event instant $t_k$ the supervisor selects one of a finite set of primitives: a pulse at $\pm\tau_{\max}$, cruise, or brake. The objective is to reach each waypoint while keeping the vehicle in the safe set $\{h \ge 0\}$ of \eqref{barrier}, where the predicted motion composes the learned models \eqref{yaw}--\eqref{long} with uncertainty $\xi$ (engagement, magnitude, $(k_v,T_v)$, and heading drift). The safety requirement is that braking from the current state keeps the vehicle in the safe set,
\begin{equation}
\label{eq:chance}
\Pr\bigl[h(p(t)) \ge 0 \ \text{for all } t \text{ on the braking maneuver}\bigr] \ge 1-\eta,
\end{equation}
\noindent with admissible risk $\eta$. Selecting among stochastic primitives over the braking horizon is a stochastic mixed-integer predictive problem, and the slew-limited weight shift makes the braking horizon, rather than the sampling period, the relevant prediction length. Instead of solving it, the supervisor uses braking as a backup policy, in the spirit of backup-set safety filters~\cite{gurriet2020scalable_acc,chen2021backup_acc}: driving continues only while the predicted braking path stays in the safe set, and the vehicle brakes otherwise. Three rules implement this design. The pulse rule \eqref{ei} compares the expected heading error of the pulse and cruise primitives over one step. The braking-path check \eqref{pathcheck} tests $h \ge 0$ along the whole brake maneuver. The worst-case stopping distance covers $(k_v,T_v)$, and the margin \eqref{margin} covers heading drift, so that the check is a sufficient condition for \eqref{chance} with $\eta = 2[1-\Phi(2)]$ per braking event under the assumptions of Proposition~\ref{prop:safety}. No barrier decay condition or optimization is enforced online; the only offline computation is a stopping-distance table from simulating \eqref{long} under the braking law over a $(\delta, v)$ grid and the $(k_v,T_v)$ corners. Replacing these rules by a short-horizon chance-constrained NMPC over the same primitives is the subject of ongoing work.

\subsection{Pulse Trigger}
After dwell, a pulse toward the goal is issued if it reduces expected heading error $e$,
\begin{equation}
\label{eq:ei}
(1-\hat{p})\,|e| + \hat{p}\,\mathbb{E}_M\bigl|\,|e| - M\bigr| < |e| - \epsilon,
\end{equation}
\noindent with engagement probability $\hat{p}$ from \eqref{probit}, lognormal magnitude $M$, and $\epsilon = 1^\circ$. Since $\mathbb{E}_M$ is insensitive to torque above threshold, the rule selects $\tau_{\max}$, and \eqref{ei} yields a heading tolerance of $23.5$--$25.5^\circ$ for $\hat{p}\in[0.28,0.64]$. It depends more on the magnitude model: about $12^\circ$ for half the median and $100^\circ$ for twice the log spread. When \eqref{ei} fails, the supervisor drives. Body rotation is opposite to flywheel torque.

\subsection{Braking and Safety Triggers}
While driving, stopping distance $\hat{d}$ is predicted from \eqref{long} (mean or worst-case over the $k_v,T_v$ box). Braking fires when along-track distance to the waypoint falls below $\hat{d}$, or when the predicted braking path leaves the safe set. With position $p$, person location $p_h$, personal radius $\rho_h$, and arena half-space margins,
\begin{equation}
\label{eq:barrier}
h(p) = \min\bigl(\|p - p_h\| - \rho,\; h_{\text{arena}}(p)\bigr),
\end{equation}
\noindent where $\rho$ inflates $\rho_h$ during braking only. With heading unit vector $n_\psi$, the braking path is the segment $p + s\,n_\psi$, $s \in [0,\hat{d}]$, and the safety trigger fires when
\begin{equation}
\label{eq:pathcheck}
h_{\text{path}} = \min_{s\in[0,\hat{d}]} h(p + s\,n_\psi) < 0 \quad\text{and}\quad h_{\text{path}} < h(p).
\end{equation}
\noindent The second condition lets a vehicle outside the safe set keep moving away from the boundary. The arena terms are affine, so their minimum lies at an endpoint, and the disk term is evaluated at the segment point closest to $p_h$; the check is exact for the straight path and, unlike a check of the stopping point alone, detects paths that cross the disk. Lateral deviation from the straight path is covered by the margin \eqref{margin}.

\subsection{Overcoming Actuator Uncertainty}
Three mechanisms tighten the learned models: worst-case $\hat{d}$; heading margin
\begin{equation}
\label{eq:margin}
\rho = \rho_h + m + 2\,\hat{d}\,\sigma_w\sqrt{\hat{d}/3},
\end{equation}
\noindent with $\sigma_w$ in rad/$\sqrt{\mathrm{m}}$ and fixed $m$; and a grid posterior on $\mu_s$ updated from first-motion detector outcomes, with $\hat{p}$ in \eqref{ei} the posterior predictive. If $\hat{p} < p_{\min}$, yaw is infeasible and the vehicle stops safely.

\begin{prop}
\label{prop:safety}
Suppose braking events are separated by at least $T_d$, $\sigma_w$ is constant, and $(k_v,T_v)$ lie in the assumed box used for $\hat{d}$. If $\rho$ satisfies \eqref{margin} with the realized braking distance, then at each braking event and at any point of the braking path, the magnitude of the lateral deviation from the straight braking path exceeds the heading term of \eqref{margin} with probability at most $2[1-\Phi(2)] \approx 4.6\%$ under the Brownian heading model. Over $N_b$ braking events per mission, a union bound gives violation probability $\le 2N_b[1-\Phi(2)]$.
\end{prop}
\noindent Proof sketch: the heading deviation at arc length $s$ is $\sigma_w W(s)$ with $W$ a Brownian motion. For small angles the lateral displacement $y(s) = \sigma_w\int_0^{s} W(r)\,dr$ is zero-mean Gaussian with variance $\sigma_w^2\int_0^{s}\!\int_0^{s} \min(r,r')\,dr\,dr' = \sigma_w^2 s^3/3 \le \sigma_w^2 \hat{d}^3/3$ for $s \in [0,\hat{d}]$. The term $2\hat{d}\sigma_w\sqrt{\hat{d}/3}$ in \eqref{margin} is therefore at least two standard deviations of $y(s)$. Deviation in either direction can reduce clearance to the arena boundaries, so the two-sided exceedance probability $2[1-\Phi(2)]$ applies per event; the union bound gives the claim. The bound holds pointwise along the path rather than uniformly over it, and it does not address discrete engagement failure; Monte Carlo rates in Section~\ref{sec:results} include all sources.

\section{Simulation Results}
\label{sec:results}
Each simulated plant draws $(\mu_s,\sigma_s)$ from the bootstrap, the log-magnitude mean from its sampling distribution, and $(k_v,T_v)$ uniformly from the uncertainty box; heading follows the random walk while translating. This in-distribution study tests whether a supervisor handles the uncertainty it was designed for, not whether the models are correct; the mismatch study probes model errors. Missions start at rest with initial heading $-35^\circ$, about $40^\circ$ offset from the bearing to the first waypoint, and visit three goals in a $5\times 4$\,m arena around a person at $\rho_h = 0.6$\,m. All designs run on the same 500 plants and noise sequences (common random numbers keyed by run index), on talc-treated wood and on a high-friction surface with $\mu_s + 0.04$\,N$\cdot$m. The nominal design follows a rigid-body model without friction: torque proportional to heading error, $|\tau| = \mu_s|e|/26^\circ$ (capped at $\tau_{\max}$), a $10^\circ$ tolerance, and direct-force stopping at 0.5\,m/s$^2$. Baselines B1 ($\tau_{\max}$ pulses, fixed $20^\circ$ tolerance, direct-force stopping) and B2 (nominal yaw, learned stopping) separate these two model errors. Ablation adds one robust mechanism at a time to the learned design; a frozen posterior and $N$-consecutive-failure abort test deterministic alternatives. Outcomes are mutually exclusive: completed safe, violation (person disk or arena exit), timeout, or safe abort.

\begin{figure}[t!]
\centering
\includegraphics[width=\columnwidth]{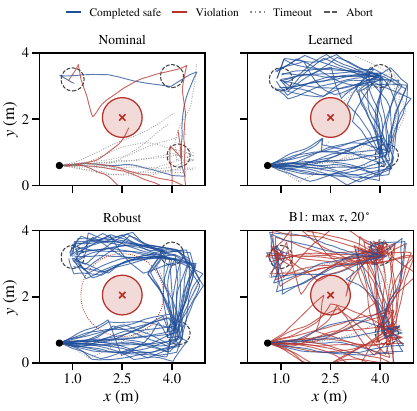}
\vspace{-6mm}
\caption{Sampled missions (25 per panel). Blue: completed safe; red: violation; gray: timeout; dashed: abort. Dotted circle: inflated keep-out for robust planning.}
\label{fig:traj}
\vspace{-3mm}
\end{figure}

\begin{table}[t!]
\caption{In-distribution Monte Carlo on talc-treated wood (500 runs, \%). Outcomes are mutually exclusive; a run may violate both disk and arena.}
\label{tab:mc}
\centering
\scriptsize
\setlength{\tabcolsep}{2.5pt}
\begin{tabular}{lccc}
\hline
 & Nominal & Learned & Robust \\
\hline
Completed safe & 1.8 & 84.0 & 92.6 \\
Violation & 20.4 & 7.2 & 1.8 \\
\quad disk / arena & 6.0 / 14.8 & 0.8 / 6.4 & 0 / 1.8 \\
Timeout / abort & 77.8 / 0 & 8.8 / 0 & 4.0 / 1.6 \\
Time safe, med.\ (s) & 210 & 125 & 125 \\
Safety brakes, mean (med.) & 0.2 (0) & 4.4 (1) & 2.5 (2) \\
Min clearance 5th pct.\ (m) & $-0.10$ & 0.05 & 0.20 \\
\hline
\multicolumn{4}{l}{High friction ($\mu_s+0.04$)} \\
Completed safe / violation & 0 / 6.8 & 1.4 / 2.0 & 0.4 / 0 \\
Timeout / abort & 93.2 / 0 & 96.6 / 0 & 5.0 / 94.6 \\
Abort time, med.\ (s) & n/a & n/a & 30.2 \\
\hline
\end{tabular}
\vspace{-2mm}
\end{table}

\begin{figure}[t!]
\centering
\includegraphics[width=\columnwidth]{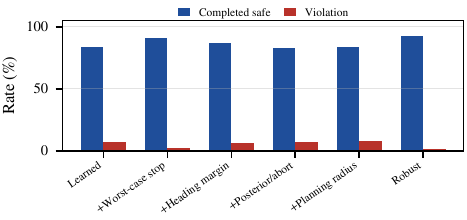}
\vspace{-5mm}
\caption{One-at-a-time ablation on talc-treated wood (500 runs). Worst-case stop alone: 91.4\% safe, 2.2\% violation.}
\label{fig:ablation}
\vspace{-3mm}
\end{figure}

\tabref{mc} reports the in-distribution results with Wilson 95\% intervals in the text. The nominal gain commands $|\tau| < \mu_s$ whenever $|e| < 26^\circ$, so 91\% of its pulses do not rotate the body, turns rarely finish (77.8\% timeout), and late direct-force braking causes its 20.4\% violations. It quantifies the cost of ignoring the friction gate rather than serving as a competitive controller; the learned design is the reference for the robust mechanisms. The learned design pulses at $\tau_{\max}$ and uses \eqref{ei}, reaching 84.0\% safe completion (80.5--87.0\%) but 7.2\% violations, mostly arena exits (mean 4.4 safety-brake events per mission, median 1, 99th percentile 82). The robust design cuts violations to 1.8\% (0.9--3.4\%) and raises 5th-percentile clearance to 0.20\,m, with 1.6\% safe aborts when the online posterior declares yaw infeasible. A false abort is an abort while the plant breakaway $\mu_s$ plus any persistent offset remains at or below $\tau_{\max}-0.01$\,N$\cdot$m (the plant still has authority); none occurred on talc-treated wood. Online posterior updates dominate the completion gap versus a frozen posterior (83.0\% with posterior/abort alone versus 93.8\% frozen; \figref{ablation}); worst-case stop alone reaches 91.4\%. McNemar tests favor robust on both completion ($p<10^{-6}$) and violations ($p=1.1\times10^{-5}$). Replacing the path check \eqref{pathcheck} by a stopping-point check changed one learned run and no robust run, since the detour planner keeps straight segments clear of the inflated disk.

\begin{figure}[t!]
\centering
\includegraphics[width=\columnwidth]{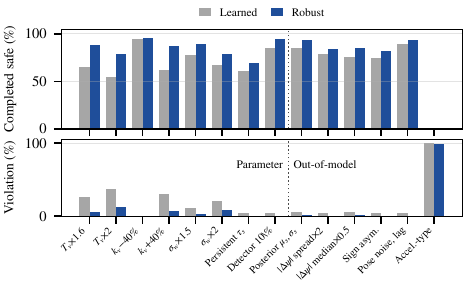}
\vspace{-5mm}
\caption{Learned versus robust under plant mismatch (300 runs per scenario). Left of the dotted line: parameters outside the design box; right: plants whose structure differs from the design model.}
\label{fig:mismatch}
\vspace{-3mm}
\end{figure}

In \figref{mismatch}, parameters outside the design box ($T_v\times 1.6$ or $\times 2$, $k_v+40\%$, $\sigma_w\times 2$) lower robust completion to 78.7--88.7\% with 6.0--12.3\% violations, against 54.3--67.0\% and 20.7--37.0\% for the learned design; a persistent breakaway offset raises robust aborts to 24\% with no violations. Among plants whose structure differs from the design model, robust completion is 93.3\% with $(\mu_s,\sigma_s)$ from the unclipped grid posterior, 85.0\% and 83.7\% with half the median rotation or twice its log spread (which four engagements cannot exclude), 82.0\% with a $\pm 0.01$\,N$\cdot$m sign-dependent breakaway (13.7\% aborts), and 93.3\% with 2\,cm pose noise and 0.1\,s latency, with at most 1.3\% violations. With both the log mean and the log spread of the magnitude drawn from their $n=4$ posterior, robust completion is 90.0\% with 0.7\% violations; because the braking and safety triggers do not use the magnitude model, magnitude errors lengthen turns rather than cause violations. An acceleration-type longitudinal response, in which speed integrates the mass offset, defeats both designs (at least 99\% violations) because the stopping tables assume \eqref{long}, so longitudinal identification is the most safety-critical open item.

\begin{figure}[t!]
\centering
\includegraphics[width=\columnwidth]{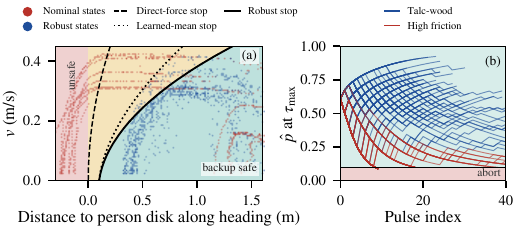}
\vspace{-6mm}
\caption{Safe sets from the learned models (100 missions). (a) Backup-safe set for states heading into the person disk, with direct-force, learned-mean, and robust stopping boundaries. (b) Yaw-feasible set: posterior engagement probability per pulse for the robust design; the shaded band triggers safe abort.}
\label{fig:safe_set}
\vspace{-3mm}
\end{figure}

\figref{safe_set} interprets these results as safe sets. In \figref{safe_set}(a), a state is backup-safe when its distance along the heading to the person disk exceeds the predicted stopping distance plus margins; the robust boundary includes the worst-case $(k_v,T_v)$ corner and the heading margin of \eqref{margin}. Among states heading into the disk, 37.3\% of nominal states lay inside it and 68.5\% lay outside the robust backup-safe set, because the direct-force model triggers braking too late. No robust state entered the disk; 8.0\% lay left of the robust boundary, which assumes cruise lead-screw position and therefore overstates the stopping distance of states already braking. In \figref{safe_set}(b), the online posterior fell below $p_{\min}$ in 93\% of high-friction missions after a median of 9 pulses and in none of the talc-treated wood missions, so the yaw-feasible set separates surfaces on which the reaction wheel retains authority from those on which it does not.

\begin{figure}[t!]
\centering
\includegraphics[width=\columnwidth]{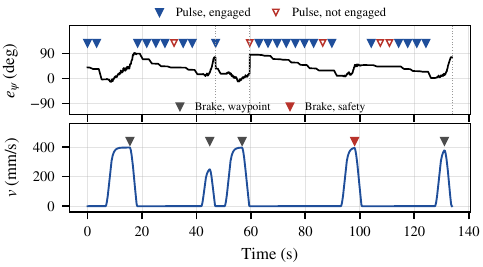}
\vspace{-6mm}
\caption{Representative robust mission on talc-treated wood. Top: heading error and pulses (filled: engaged). Bottom: speed and brakes; dotted lines mark waypoint switches.}
\label{fig:timeline}
\vspace{-4mm}
\end{figure}

\figref{timeline} shows a representative robust mission: waypoint switches (dotted), engaged and failed pulses, and goal and safety brakes. On high friction the robust supervisor aborts safely (94.6\% abort, 0\% violation, median 30.2\,s) when $\hat{p}$ collapses. Baseline B1 (max $\tau$, $20^\circ$ tolerance) reaches 28.6\% completion with 67.2\% violations; B2 (nominal yaw, learned stop) reaches 5.4\%.

These results carry several limitations. The yaw model rests on 24 ordered pulses on one surface, of which 19 were stationary and 4 engaged, so an order or sign offset of about 0.03\,N$\cdot$m cannot be excluded, the threshold sharpness is poorly constrained, and the magnitude distribution is weakly identified. The longitudinal lag and response type are assumed, and $\sigma_w$ came from five trials whose heading change followed the commanded torque. The Monte Carlo rates are therefore conditional on these models; they compare designs under a common uncertainty model rather than predict hardware performance. All closed-loop outcomes come from simulation; the out-of-model plants are themselves hypotheses, and the supervisor is not compared with an optimal predictive policy. Closing these gaps requires randomized-order sweeps over torque magnitude, sign, and pulse duration on several surfaces, logged longitudinal step and braking responses, pulse-free translating trials for $\sigma_w$, and closed-loop hardware experiments.

\section{Conclusions}
\label{sec:conclusions}
We presented human-inspired actuation of a black-box balancing wheel, in which a weight shift produced longitudinal motion and a momentum-limited reaction wheel produced yaw, together with data-driven models that captured the friction-gated engagement of the reaction wheel and the slew-limited response of the weight shift. Building on these models, an uncertainty-aware hybrid supervisor with a backup-safe braking policy combined probabilistic pulse scheduling, braking-path safety checks, and an online posterior on the breakaway threshold. In simulation on plants drawn from the fitted models, the robust supervisor reached 92.6\% safe completion with 1.8\% violations on talc-treated wood, compared with 84.0\% and 7.2\% for the learned design, and it aborted safely on high friction with no violations, which showed that accounting for actuator uncertainty in both triggering and safety tightening was essential on a closed platform whose balance controller could not be commanded.

\section*{Acknowledgments}
\noindent We thank Nokia Bell Labs for a research award and Mike Coss, Head of Robotics Research, for his support.

\begin{small}

\end{small}

\end{document}